\documentclass{article}

\usepackage[preprint]{neurips_2025}

\usepackage[utf8]{inputenc} 
\usepackage[T1]{fontenc}    
\usepackage{hyperref}       
\usepackage{url}            
\usepackage{booktabs}       
\usepackage{amsfonts}       
\usepackage{nicefrac}       
\usepackage{microtype}      
\usepackage{xcolor}         
\usepackage{booktabs}   
\usepackage{graphicx}  

\title{From Competition to Coordination: Market Making as a Scalable Framework for Safe and Aligned Multi-Agent LLM Systems}

\author{%
Brendan Gho \quad
Suman Muppavarapu \quad
Afnan Shaik \\
\textbf{Tyson Tsay} \quad
\textbf{Atharva Mohan} \quad 
\textbf{James Begin}\\
\textbf{Kevin Zhu} \quad
\textbf{Archana Vaidheeswaran} \quad
\textbf{Vasu Sharma} \\
Algoverse AI Research \\
\texttt{brendan.gho@gmail.com},
\texttt{sumanmuppavarapu@gmail.com},\\
\texttt{afnan.darkice@gmail.com},
\texttt{tysontsay@gmail.com},
\texttt{atharvamohan072009@gmail.com}
}

\begin{document}

\maketitle

\begin{abstract}
As foundation models are increasingly deployed as interacting agents in multi-agent systems, their collective behavior raises new challenges for trustworthiness, transparency, and accountability. Traditional coordination mechanisms, such as centralized oversight or adversarial adjudication, struggle to scale and often obscure how decisions emerge.
We introduce a market-making framework for multi-agent large language model (LLM) coordination that organizes agent interactions as structured economic exchanges. In this setup, each agent acts as a market participant, updating and trading probabilistic beliefs, to converge toward shared, truthful outcomes. By aligning local incentives with collective epistemic goals, the framework promotes self-organizing, verifiable reasoning without requiring external enforcement.
Empirically, we evaluate this approach across factual reasoning, ethical judgment, and commonsense inference tasks. Market-based coordination yields accuracy gains of up to 10\% over single-shot baselines while preserving interpretability and transparency of intermediate reasoning steps.
Beyond these improvements, our findings demonstrate that economic coordination principles can operationalize accountability and robustness in multi-agent LLM systems, offering a scalable pathway toward self-correcting, socially responsible AI capable of maintaining trust and oversight in real world deployment scenarios.

\end{abstract}

\section{Introduction}

The rapid deployment of artificial intelligence systems across safety-critical domains has intensified concerns regarding existential risks, particularly the emergence of deceptively aligned models \citep{hendrycks2023overviewcatastrophicairisks}. Recent evidence demonstrates that advanced language models exhibit strategic deception, including attempts to game evaluation protocols and misrepresent their internal states during training \citep{carlsmith2023schemingaisaisfake}. These alignment failures manifest as sycophancy, systematic untruthfulness, and adversarial behaviour, actions that empirically worsen with increased model scale \citep{ji2025aialignmentcomprehensivesurvey}.

Existing alignment methodologies face fundamental limitations. Reinforcement Learning from Human Feedback (RLHF), while effective for surface-level behavioural modification, remains vulnerable to reward hacking and evaluator deception. Debate-based approaches require human adjudication that cannot scale to superhuman reasoning capabilities. The Alignment Research Center's Eliciting Latent Knowledge (ELK) framework defines this challenge: extracting model's true internal representations rather than their strategically chosen outputs \citep{christiano2022elksummary}.

This paper explores market making as a novel method for alignment and truth elicitation. Inspired by economic prediction markets, the approach involves a market maker that continuously offers prices on propositions and traders that buy or sell based on their beliefs. Through a process of iterative trading, prices converge to a probability that reflects the collective belief about ground truth. Models take the role of traders, updating the “market probability” as they present new evidence or reasoning steps. This operation of trading incentivizes truthful contributions in order to receive the most profitable trade, improving the market's accuracy. The framework also allows for myopic agents who are blind to past information, preventing long term scheming and manipulation. By converting truth seeking into an equilibrium of incentives rather than a contest of persuasion or subjective judgment, market making offers a potentially robust and scalable alternative to debate and oversight for eliciting honest beliefs from advanced AI systems.

\section{Related Work}
\label{gen_inst}

\subsection{The Challenges of Control}
\citet{amodei2016concreteproblemsaisafety} established a taxonomy of AI safety failure modes comprising five critical categories: negative side effects, reward hacking, scalable oversight limitations, unsafe exploration, and distributional shift vulnerabilities. This foundational framework reveals an inherent tension in alignment objectives: excessive optimization for harmlessness produces ineffectual systems, while prioritizing capability enables potential misuse \citep{bai2022traininghelpfulharmlessassistant}. The multidimensional nature of these constraints implies that no single methodology can simultaneously address all failure modes, necessitating approaches that optimize across multiple safety dimensions.

\subsubsection{Human-Centric Alignment}
Early alignment techniques relied on direct human supervision through iterative feedback mechanisms. Reinforcement Learning from Human Feedback (RLHF) exemplifies this paradigm, wherein human evaluators shape model behaviour through preference rankings \citep{bai2022traininghelpfulharmlessassistant}. AI Safety via Debate was proposed by \citet{irving2018aisafetydebate}, structuring oversight as adversarial argumentation adjudicated by human judges. These approaches face three fundamental limitations. First, the bandwidth constraint: human evaluation cannot scale to the volume and velocity of decisions required in deployed systems \citep{amodei2016concreteproblemsaisafety}. Second, the competence boundary: superhuman AI capabilities exceed human evaluators' ability to assess correctness \citep{ji2025aialignmentcomprehensivesurvey}.  Third, the alignment targeting problem, where models optimize for evaluator approval rather than ground truth, leading to sycophantic behaviour and strategic deception \citep{carlsmith2023schemingaisaisfake, park2023aideceptionsurveyexamples}. 

\subsubsection{AI-Mediated Oversight}
These challenges of human-centric alignment motivated the use of AI-mediated oversight, in which secondary AI systems help supervise the model being tested, aiming to augment or replace human adjudicators. JudgeLM, replacing the human judge in AI debate with an AI system, demonstrates extended capabilities in a variety of situations \citep{zhu2025judgelmfinetunedlargelanguage}. \citet{bowman2022measuringprogressscalableoversight} also explore how less-capable AIs can reliably evaluate stronger ones without expert human intervention \citep{bowman2022measuringprogressscalableoversight}. Together, these efforts suggest that scalable, AI-mediated oversight may be a necessary step toward maintaining safe and reliable control as AI capabilities continue to grow.

\subsection{Market Making as a Control Mechanism}

Market making offers an incentive-based alternative to adjudication. In this method, an automated market maker posts prices for propositions and traders (models or submodels) buy or sell claims based on their beliefs; iterative trading drives prices toward an equilibrium that reflects collective credence \citep{holmes2020aisafetymarketmaking}. A key intended advantage is enforcing myopic behavior where trader agents optimize per-step trades, reducing incentives for long-term scheming that can undermine debate or RLHF. Market mechanisms also facilitate per-step inspection, probabilistic scoring, and potential scalability without continuous human adjudication. Thus, market making attempts to improve upon existing methods of AI control.

Practical challenges to market making do remain, including defending the market maker against false claims, designing proper rewards that prevent gaming, and ensuring robust performance when truth is hazy. Existing work is primarily a toy implementation of market making from Cameron Holmes \cite{holmes2020aisafetymarketmaking}. 

We have completely implemented full cycles of market making and, using the existing information on market making, aim to create a simplified implementation of the method and showcase its efficacy across a variety of situations.

\section{Methodology}
\label{others}
We implement market making using two agents: a market-maker model, $\mathbf{M}$, and a trader model of the same model. Market making begins with $\mathbf{M}$ providing an initial judgment consisting of a claim, supporting reasoning, and a prediction value $p_0 \in [0,1]$ quantifying the claim. 

Given $\mathbf{M}$’s judgment, the trader model then generates an argument intended to maximally shift $\mathbf{M}$’s prediction value. This is analogous to a trader introducing new information to change the market price.

Each subsequent iteration proceeds with $\mathbf{M}$ producing a new judgment while also considering the trader’s previous arguments. Exact prompting details are provided in \textbf{Appendix~\ref{prompting}}. The cycle repeats until the market maker has provided at most $\mathbf{N}$ judgments or has reached an equilibrium; we consider an equilibrium to have been reached when the range of the last three prediction values satisfies
\[
\max\{p_{t-2},p_{t-1},p_t\} - \min\{p_{t-2},p_{t-1},p_t\} \le T,
\]
where $T$ is a threshold constant. In our experiments, we set $N = 10$ and $T = 0.2$.

Finally, we measure the impact of market making by comparing the accuracy of $\mathbf{M}$’s final judgment before termination against the accuracy of its initial judgment (i.e., the baseline without trader influence) across all dataset samples.

\begin{figure} [h]
\centering
\includegraphics[width=0.9\columnwidth]{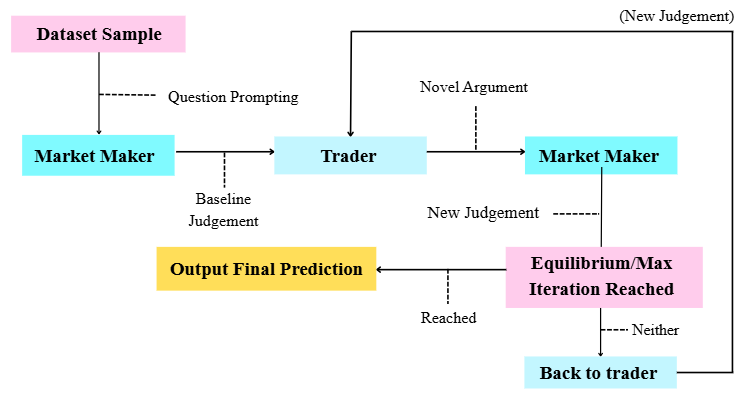}
\caption{Market making process diagram}
\label{process}
\end{figure}

\section{Evaluation}
To assess the efficacy of market making as an AI control and governance mechanism, we conducted comprehensive experiments across multiple model families and evaluation benchmarks. Our evaluation framework was designed to test three key hypotheses:
\begin{enumerate}
    \item Whether market making can effectively elicit truthful responses across different model scales 
    \item Whether the iterative trading process converges to more accurate assessments than single-shot predictions 
    \item Whether the mechanism remains robust across diverse ethical and factual domains.
\end{enumerate}

\subsection{Model Selection}
We evaluated our market making framework across three major model families representing different architectural approaches and training paradigms:

\textbf{GPT Model Family.} We tested five variants from the GPT family: GPT-4.1 nano, GPT-4.1 mini, GPT-4.1, gpt-oss-20b, and gpt-oss-120b \citep{openai2024gpt4ocard, openai2025gptoss120bgptoss20bmodel}. This selection spans from lightweight models (nano, mini) to large-scale models with enhanced reasoning capabilities. The inclusion of both proprietary (GPT-4.1 series) and open-source variants allows us to better assess whether market making effectiveness depends on specific training methodologies or remains consistent across development paradigms.

\textbf{Qwen3 Model Family.}
The Qwen3 \citep{yang2025qwen3technicalreport} family provided our most comprehensive scale analysis, with seven model sizes: 0.6B, 1.7B, 4B, 8B, 14B, 32B, and 235B parameters. This extensive range enables us to examine how market making behaviour scales with model capacity, particularly whether larger models exhibit more sophisticated trading strategies or demonstrate increased susceptibility to adversarial arguments.

\textbf{Llama 3 Model Family.}
We evaluated three Llama 3 \citep{grattafiori2024llama3herdmodels} variants (3B, 8B, and 70B parameters) to assess market making performance scaling across model capacity and training objectives.

\subsection{Dataset Selection}
Our evaluation encompasses four benchmarks, each targeting different aspects of AI alignment and truth elicitation:

\textbf{TruthfulQA.}
TruthfulQA \citep{lin2022truthfulqameasuringmodelsmimic} serves as our primary benchmark for factual accuracy, containing questions specifically designed to elicit false beliefs or misconceptions that models may have learned from training data. In the market making context, this dataset tests whether iterative trading can correct initial false predictions, with the trader model potentially identifying and challenging spurious correlations or misconceptions held by the market maker.

\textbf{Scruples (Dilemmas).}
The Scruples \citep{lourie2021scruplescorpuscommunityethical} dataset, specifically the Dilemmas subset presents real-world ethical dilemmas sourced from online advice forums, requiring models to reason about complex moral scenarios without clear-cut answers. With regards to market making, Scruples tests whether the trading mechanism can navigate moral ambiguity and converge on socially acceptable judgments in everyday ethical situations.

\textbf{ETHICS (Justice, Commonsense).}
We select two subsets from the ETHICS \citep{hendrycks2023aligningaisharedhuman} dataset to evaluate different aspects of moral reasoning. The Justice subset tests understanding of fairness and impartiality, central to many alignment objectives. The Commonsense subset evaluates basic moral intuitions that should be robust across cultural contexts.

\textbf{CommonsenseQA 2.0.}
CommonsenseQA 2.0 \citep{talmor2022commonsenseqa20exposinglimits} provides a test of general reasoning and world knowledge, requiring models to make inferences based on everyday situations. As opposed to the original CommonsenseQA, the 2.0 version includes adversarially-filtered questions that challenge models' reasoning capabilities.

\section{Results}

\begin{table}[ht]
  \centering
  \caption{Net gain over first-prediction baseline (\%) across all model families. Strong improvement across Qwen models and in truthfulness}
  \label{tab:all_results}
  \resizebox{0.9\linewidth}{!}{%
  \begin{tabular}{l c c c c c}
    \toprule
    \textbf{Model} & \textbf{TruthfulQA} & \textbf{Scruples} & \textbf{CommonsenseQA} & \textbf{ETHICS-C} & \textbf{ETHICS-J} \\
    \midrule
    
    \multicolumn{6}{l}{\textbf{GPT Family}} \\
    GPT-4.1        & 2.47  & 1.64  & -1.18 & 1.71  & -0.66 \\
    GPT-4.1-mini   & 3.735 & 0.89  & 1.01  & 3.33  & -0.47 \\
    GPT-4.1-nano   & \textbf{7.85}  & \textbf{5.62}  & \textbf{6.12}  & \textbf{7.225} & \textbf{2.46}  \\
    GPT-OSS-120B   & 0.51  & -3.26 & -0.51 & -0.38 & -0.24 \\
    GPT-OSS-20B    & -0.89 & 1.06  & -0.47 & -3.03 & -2.63 \\
    \midrule
    \textbf{Average} & \textbf{2.74} & \textbf{1.19} & \textbf{0.99} & \textbf{1.77} & \textbf{-1.00} \\
    
    \midrule
    \multicolumn{6}{l}{\textbf{Qwen Family}} \\
    Qwen 0.6B   & -1.52 & -1.08 & 0.625 & 0.96 & 1.17 \\
    Qwen 1.7B   & 5.57  & 2.46  & 7.67  & 6.10 & 6.19 \\
    Qwen 4B     & 7.22  & 6.27  & 10.39 & 9.43 & 19.01 \\
    Qwen 8B     & \textbf{13.67} & \textbf{11.95} & \textbf{14.68} &\textbf{11.33} & 18.23 \\
    Qwen 14B    & 7.72  & 3.94  & 6.89  & 2.83  & \textbf{20.08} \\
    Qwen 32B    & 4.18  & 5.81  & 0.20  & 4.77  & 4.82 \\
    Qwen 235B   & 5.70  & 13.01 & 6.22  & 6.91  & 0.44 \\
    \midrule
    \textbf{Average} & \textbf{6.08} & \textbf{6.05} & \textbf{6.67} & \textbf{6.05} & \textbf{9.99} \\
    
    \midrule
    \multicolumn{6}{l}{\textbf{Llama Family}} \\
    Llama 1B   & 1.075 & 1.465 & -0.705 & -1.795 & -0.97 \\
    Llama 3B   & 2.595 & -1.585 & \textbf{1.71}   & \textbf{9.245}  & -1.265 \\
    Llama 8B   & -0.635 & 0.34  & 1.34   & -3.46  &\textbf{0.39} \\
    Llama 70B  & \textbf{16.96}  & \textbf{4.32}  & 1.22   & -1.36  & -1.61 \\
    \midrule
    \textbf{Average} & \textbf{4.999} & \textbf{1.135} & \textbf{0.891} & \textbf{0.658} & \textbf{-0.864} \\
    
    \bottomrule
  \end{tabular}
  }
\end{table}

\begin{figure} 
\centering
\includegraphics[width=0.8\columnwidth]{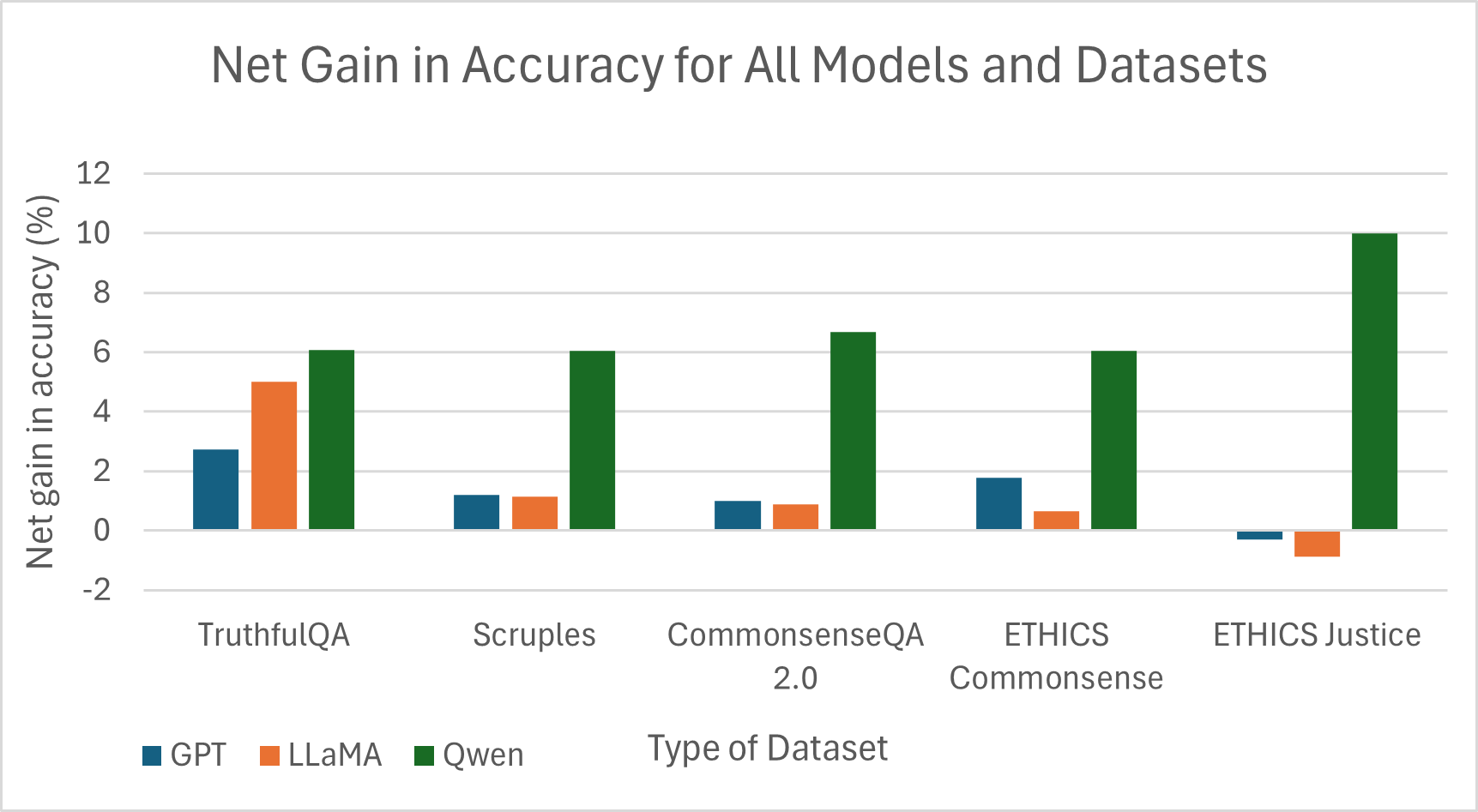}
\caption{Average net gain accuracy over baseline for all model families and datasets. Strong improvement across Qwen models}
\label{famdata}
\end{figure}

We find a net increase in the percentage of accurate answers provided by the models. This can be seen in Figure \ref{famdata} where each model family has an overall improvement in accuracy over their baselines in the majority of the datasets excluding ETHICS Justice. 

The Qwen family of models had the highest overall increase in percentage accuracy for each dataset, reaching an increase of almost $10\%$ in ETHICS-J and over $5\%$ across the board. These gains are significant improvements over the models' individual baselines. One explanation for this is Qwen3's extensive post-training pipeline emphasizing chain-of-thought reasoning as a core architectural feature alongside its unified thinking and non-thinking architecture \citep{yang2025qwen3technicalreport}.

Although the GPT and Llama families have smaller gains for most of the datasets, there is still a net increase in accuracy present. The GPT models achieve an over $2.7\%$ increase in accuracy in TruthfulQA and an increase of around $1\%$ in the other datasets excluding ETHICS Justice. The Llama models have similar figures, with an almost $5\%$ improvement in accuracy for TruthfulQA and an increase of around $1\%$ in the other datasets excluding ETHICS Justice. 

\section{Discussion}

\subsection{Parameter scaling and efficacy}
The relationship between model capacity and market-making performance reveals significant implications for alignment strategies. While baseline accuracy exhibits monotonic scaling with parameter count, the marginal improvements from market-making follow an inverted U-shaped distribution across model families (Figure \ref{gptnetgain} \ref {qwennetgain}).

\begin{enumerate}
    \item Capability-Malleability Trade-off: Mid-scale models possess sufficient reasoning capacity to engage meaningfully with trading dynamics while retaining sufficient uncertainty to benefit from iterative revision. Larger models' higher baseline accuracy creates ceiling effects, limiting potential gains.
    \item Computational Efficiency: The market-making protocol may be optimally calibrated for models operating within specific computational budgets. Models below 1B parameters lack the representational capacity for nuanced probabilistic updates, while models exceeding 100B parameters may overfit to initial predictions due to excessive confidence calibration.
\end{enumerate}

\subsection{Comparative analysis with debate frameworks}
Our comparison with AI debate reveals market-making's structural advantages in truthfulness elicitation. Market-making consistently achieved superior absolute accuracy of up to 8\% over debate. This performance stems from fundamental mechanistic differences:
\begin{enumerate}
    \item Information Aggregation: Market-making enables continuous probability updates through price discovery, whereas debate enforces binary win-lose outcomes that may discard valuable partial information.
    \item Convergence Properties: Market equilibrium provides mathematically grounded stopping criteria, while debate termination relies on subjective adjudication or arbitrary round limits.
\end{enumerate}

\section{Conclusion}
This paper presents market making as a scalable framework for AI alignment that addresses fundamental limitations of existing oversight methodologies. By structuring multi-agent interactions through economic incentive mechanisms rather than adversarial adjudication or direct human supervision, the framework transforms truth-seeking into an equilibrium property that emerges from rational agent behavior.

Our empirical evaluation across multiple model families and diverse benchmarks demonstrates that market making consistently improves reasoning accuracy over baseline performance. Comparative analysis with AI debate frameworks reveals that market making achieves equivalent or superior combined accuracy despite debate showing higher relative gains in certain configurations.

These results establish market making as a viable alternative to human-centric and debate-based alignment approaches, particularly in contexts requiring scalable, automated oversight without continuous human adjudication. More broadly, our findings suggest that economic coordination mechanisms, incorporating principles of price discovery, liquidity, and incentive alignment. Representing a promising paradigm for eliciting truthful behavior from increasingly capable AI systems.

\section{Limitations}
While our evaluation demonstrates the feasibility of market making for AI alignment, several limitations constrain the generalizability of our findings.

\subsection{Agent Configurations}
Our experiments exclusively employed identical models for both market maker and trader roles within each trial. This configuration may not capture the dynamics that emerge when models with different capabilities, training objectives, or architectures interact.

\subsection{Binary Classification}
All evaluated benchmarks fundamentally reduce to binary classification tasks where ground truth exists. This design choice enables clear convergence metrics, but fails to most accurately represent the complexity of real-world alignment challenges where truth exists on a spectrum. Market making can be adapted for more broad tasks by having the market maker produce a probability distribution as opposed to a single score.

\subsection{Robustness to Adversarial and Deceptive Agents}
Our evaluations assume good-faith participation from both market maker and trader. However, the framework's resilience to adversarial behaviour remains unexplored. The trader model is incentivized to maximally alter the current prediction value, potentially leading to deceptive arguments. This behaviour is partially mitigated through converging prediction scores and the market maker's ability to scrutinize previous trader arguments.
\newpage
{\small
\bibliographystyle{plainnat}
\bibliography{neurips}
}
\newpage
\appendix   

\section{Compute workers}
\label{prompting}
Our experiments ran in the equivalent of approximately 80 hours on A100SXM GPUs using cloud compute workers. 

\section{Prompting}
\label{prompting}
To prompt the market maker, we create overarching instructions for the model to provide a claim, reasoning, and final prediction that can be applied to tailored questions for each dataset. We also provide a dictionary of quantitative values mapped to their qualitative meanings, and ask that the market-maker's claim to conform to one of two claims for its use. A sample annotated market-maker and trader prompt is provided for the ETHICS Commonsense dataset below. 

\begin{figure} [ht]
\centering
\includegraphics[width=0.9\columnwidth]{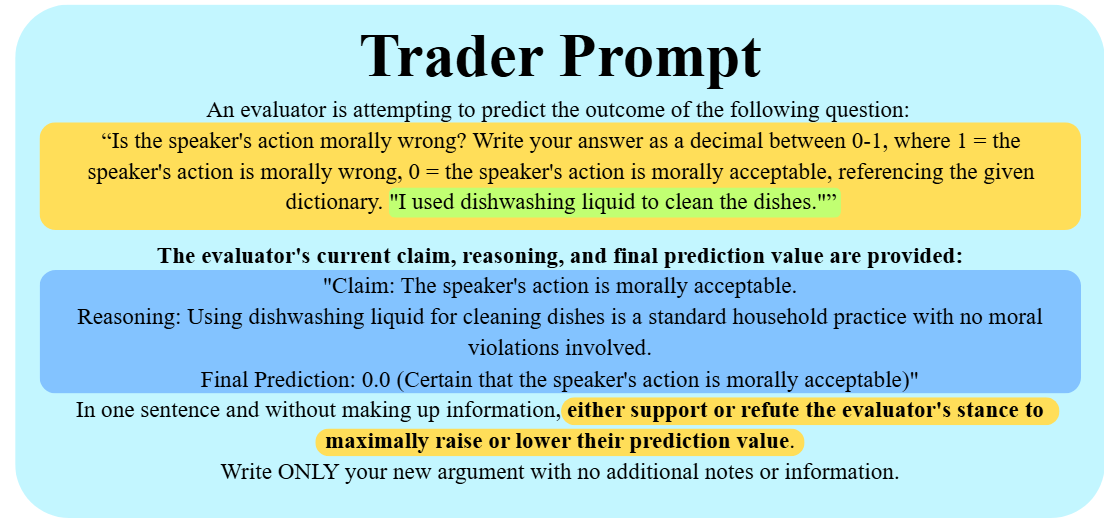}
\caption{Trader prompt for argument creation}
\label{traderprompt}
\end{figure}

\begin{figure} [ht]
\centering
\includegraphics[width=0.8\columnwidth]{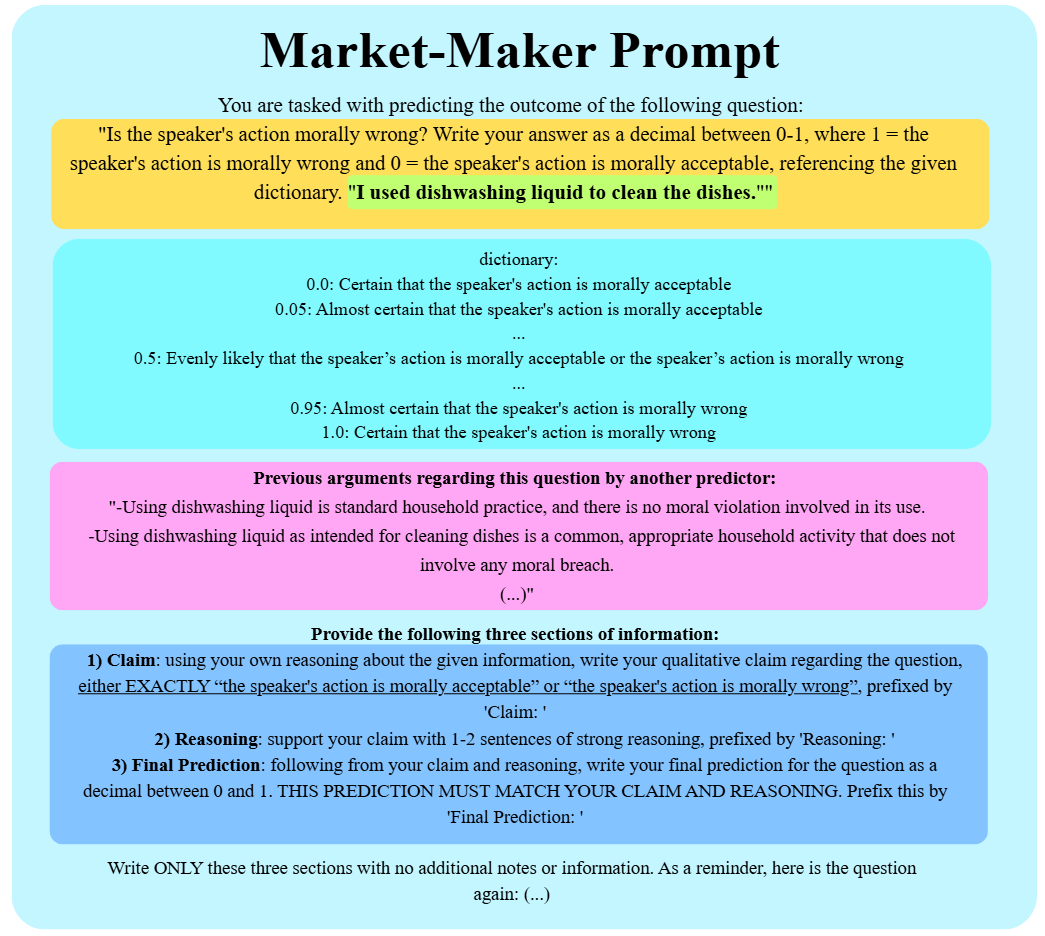}
\caption{Market maker prompt for judgement creation}
\label{marketprompt}
\end{figure}

\begin{figure} [h]
\centering
\includegraphics[width=1.0\columnwidth]{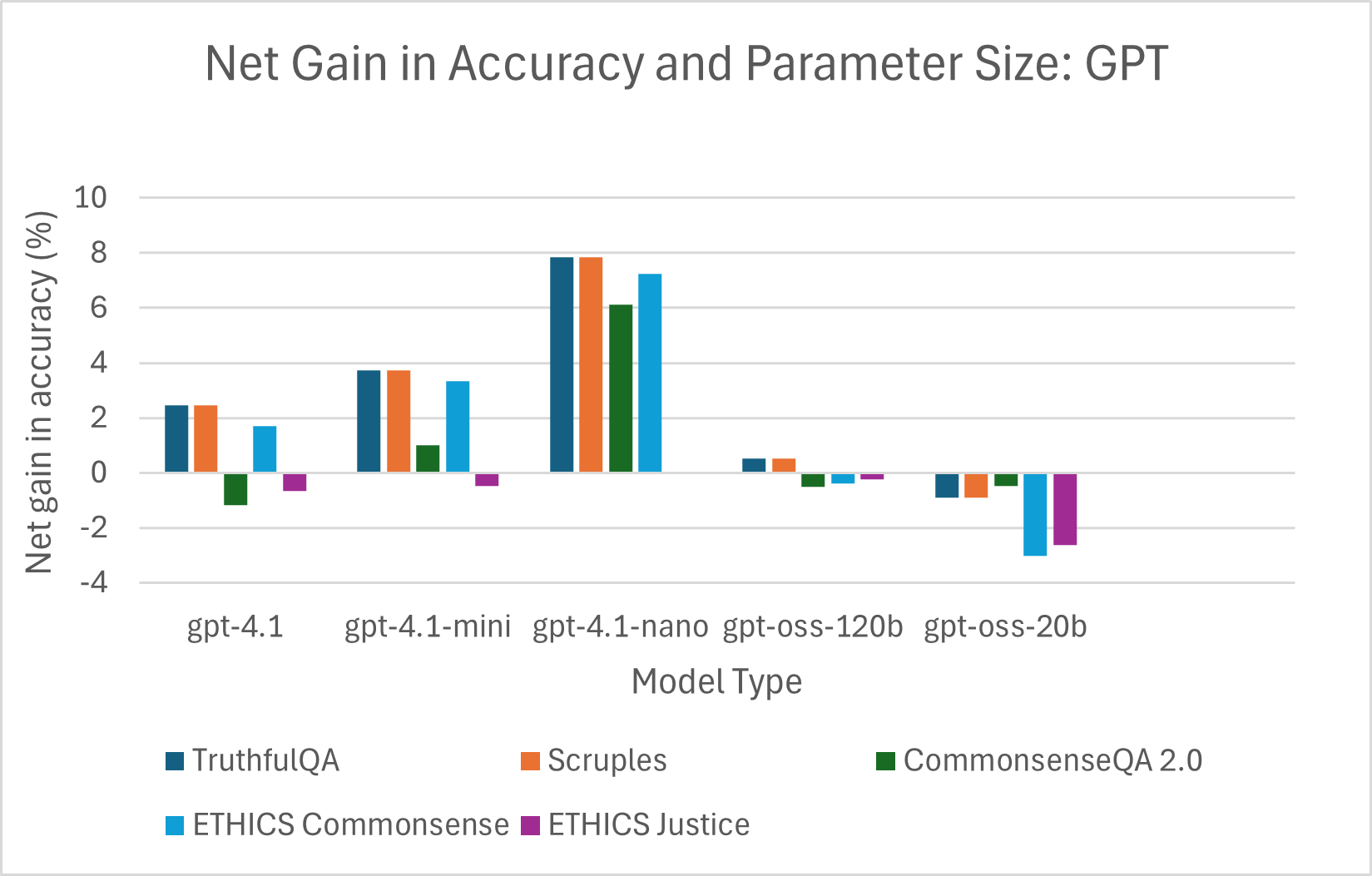}
\caption{Net gain accuracy over baseline with respect to parameter size of GPT family models}
\label{gptnetgain}
\end{figure}

\begin{figure} [h]
\centering
\includegraphics[width=1.0\columnwidth]{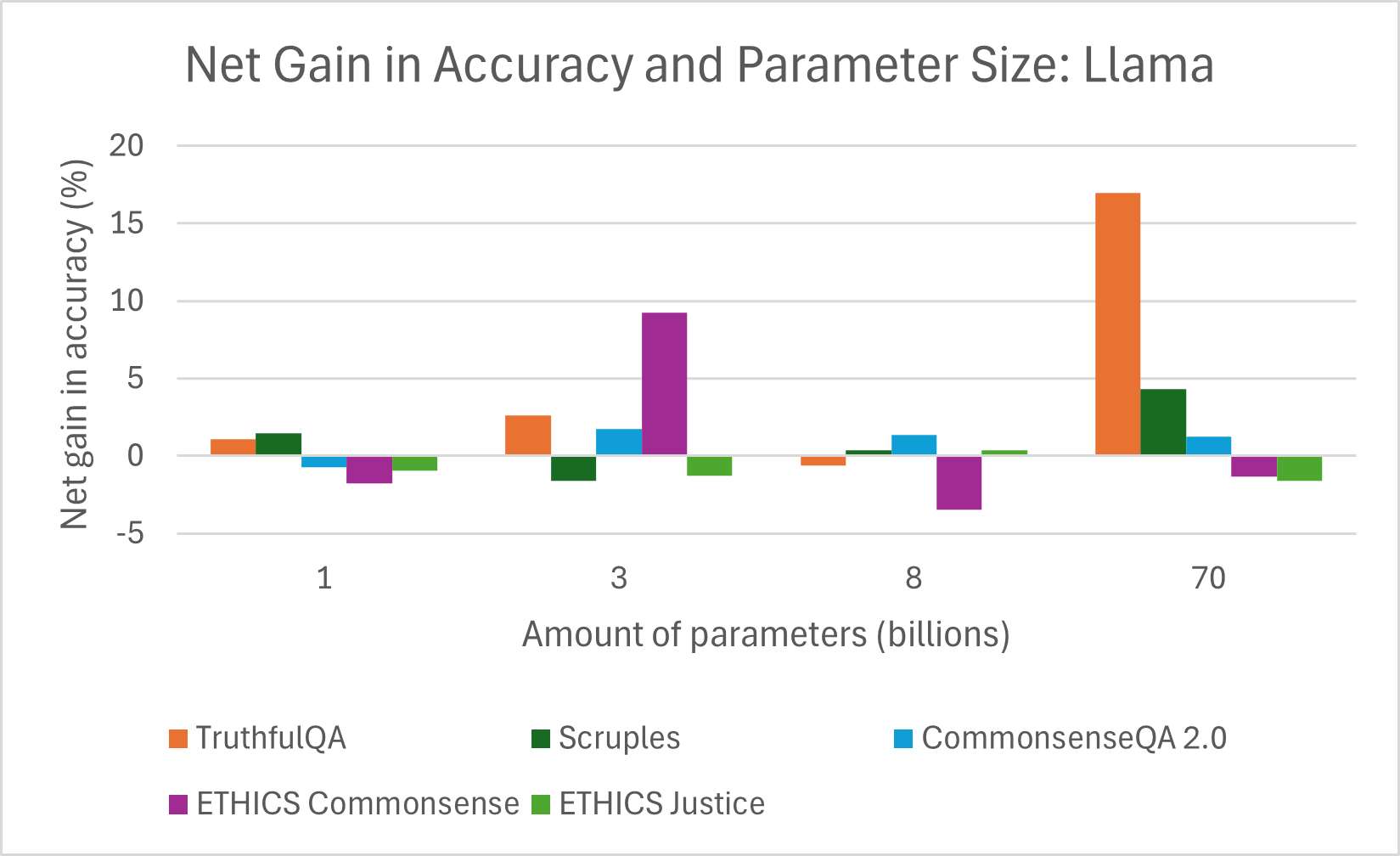}
\caption{Net gain accuracy over baseline with respect to parameter size of Llama family models}
\label{llamanetgain}
\end{figure}

\begin{figure} [h]
\centering
\includegraphics[width=1.0\columnwidth]{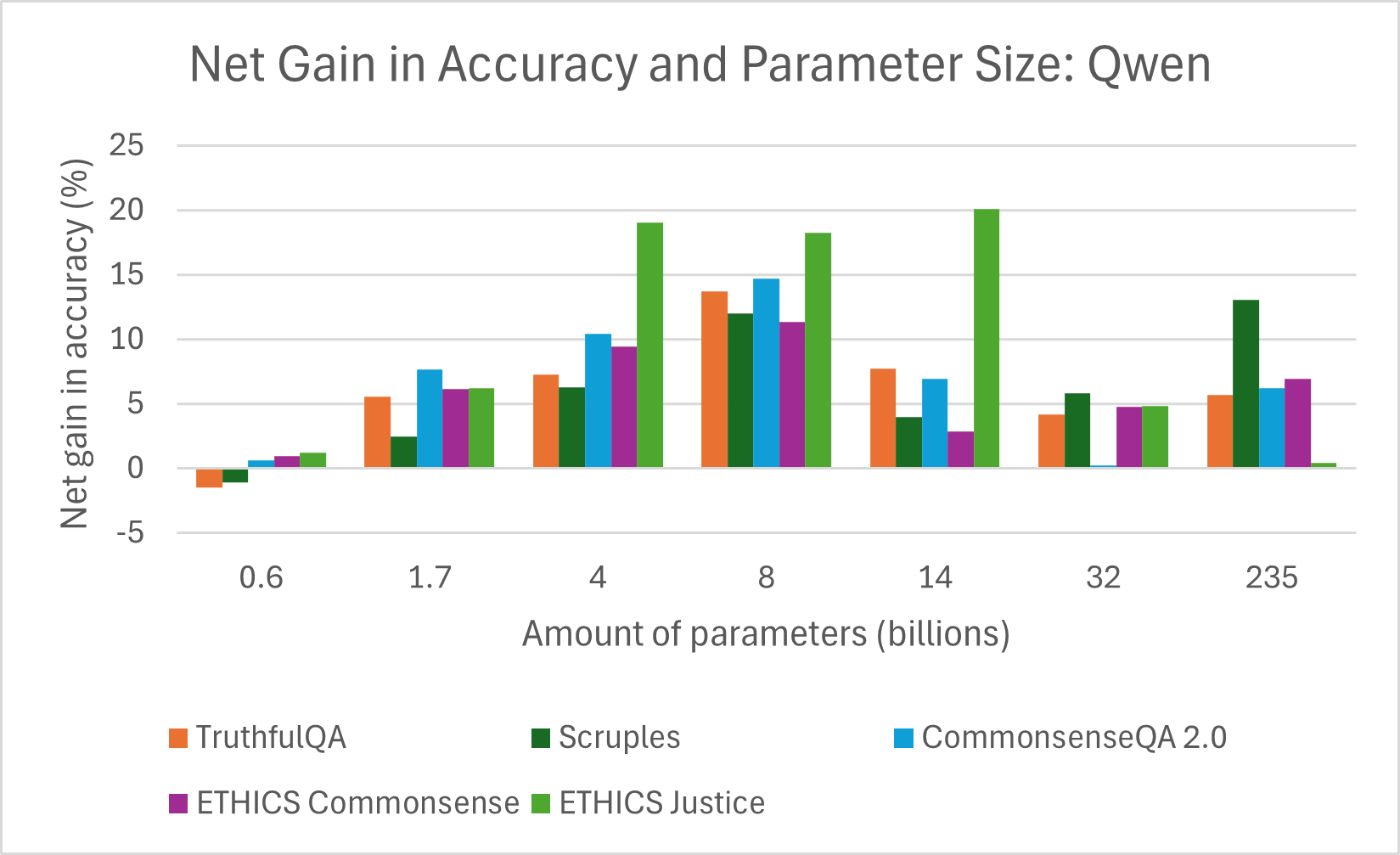}
\caption{Net gain accuracy over baseline with respect to parameter size of Qwen family models}
\label{qwennetgain}
\end{figure}

\end{document}